\documentclass[aps,floatfix,prl,twocolumn,a4paper,10pt,citeautoscript,
			   preprintnumbers,superscriptaddress,showpacs]{revtex4-1}

\usepackage[english]{babel}	

\usepackage{amsmath,amssymb,amsfonts} 
\usepackage{bbm}		

\usepackage{graphicx}	
\usepackage{graphics}	
\usepackage{xcolor}		

\DeclareGraphicsExtensions{.pdf}	
\usepackage[colorlinks=true, pdfstartview=FitV,linkcolor=blue, 
	citecolor=blue, urlcolor=blue]{hyperref}

\makeatletter
\newcommand*{\Equation}{\@ifstar\sEquation\oEquation}
\newcommand{\sEquation}[1]{\begin{equation*}#1\end{equation*}}
\newcommand{\oEquation}[2]{  \begin{equation}\label{#1}#2\end{equation} }
\makeatother

\newcommand{\bs}{\boldsymbol} 

\newcommand{\Figref}[1]{Fig.~\ref{#1}}
\newcommand{\Eqref}[1]{\eqref{#1}}

\newcommand{\groupU}[1]{U(#1)}   
\newcommand{\groupSU}[1]{SU(#1)} 
\newcommand{\groupZ}[1]{\mathbb{Z}_{#1}} 

\newcommand{\Exp}[1]{\mathrm{e}^{#1}}

\renewcommand\Im{\mathrm{Im}}

\newcommand{\Grad}{\bs \nabla}

\newcommand{\Curl}{\bs \nabla\times}

\newcommand{\x}{\bs x}

\newcommand{\ez}{\bs e_z}

\newcommand{\D}{{\bs D}}
\newcommand{\A}{{\bs A}}
\newcommand{\B}{{\bs B}}
\newcommand{\J}{{\bs J}}
\newcommand{\F}{\mathcal{F}}
\newcommand{\G}{\mathcal{G}}

\begin{document}
\title{Vortex matter in  
\texorpdfstring{$U(1)\times U(1)\times\mathbb{Z}_2$}{U(1)xU(1)xZ2} 
phase-separated superconducting condensates}

\author{Julien~Garaud}
\affiliation{Department of Theoretical Physics, KTH-Royal Institute of Technology, Stockholm, SE-10691 Sweden}
\affiliation{Department of Physics, University of Massachusetts Amherst, MA 01003 USA }
\email{garaud.phys@gmail.com}
\author{Egor~Babaev}
\affiliation{Department of Theoretical Physics, KTH-Royal Institute of Technology, Stockholm, SE-10691 Sweden}
\date{\today}

\begin{abstract}

We study the properties of vortex solutions and magnetic response 
of two-component $U(1)\times U(1)\times\mathbb{Z}_2$ 
superconductors, with phase separation driven by intercomponent 
density-density interaction. Such a theory can be viewed arising 
from the breakdown of $SU(2)$ symmetry by a biquadratic 
interaction between the components of the field. Depending on the 
symmetry-breaking term, there are two ground-state phases: one where 
both components of the doublet are equal (the miscible phase) and one 
where only one component assumes a non zero vacuum expectation value 
(the immiscible state). In the latter phase, the spectrum of topological 
excitations contains both domain walls and vortices. 
We show the existence of another kind of excitation that has properties 
of both topological excitations at the same time. They combine
vorticity together with a circular domain wall, interpolating 
between inequivalent broken states, that shows up as a ring of localized 
magnetic flux. 
Asymptotically, this resembles a vortex carrying multiple flux quanta, 
but because the magnetic field is localized at a given distance from 
the center this looks like a pipe. The isolated multiquanta pipelike 
vortices can be either stable or metastable, even if the system is not 
type-1. We also discuss the response of such a system to an externally 
applied magnetic field.

\end{abstract}

\maketitle


In the recent years, there has been growing interest in models of 
superconductivity described by more than one superconducting condensate. 
This interest  follows from the growing number of known materials that 
are described by multiple condensates. One could mention multi-band 
superconductors such as MgB$_2$ \cite{Nagamatsu.Nakagawa.ea:01} or
iron based superconductors \cite{Chu.Koshelev.ea:09}. Also multicomponent 
models apply to describe unconventional superconductors such as Sr$_2$RuO$_4$ 
that is an exotic superconductor with chiral $p_x+ip_y$ pairing symmetry 
\cite{Maeno.Hashimoto.ea:94,Mackenzie.Maeno:03}, or heavy fermion compounds
such as UPt$_3$ \cite{Joynt.Taillefer:02}.

The macroscopic physics of multicomponent superconductors is described 
by Ginzburg-Landau free energy with multiple condensates, that is, 
a field theory of multiple complex scalars charged under the same 
$\groupU{1}$ gauge field. 
There, new physics that has no counterpart in single component 
systems arises. This comprises vortices carrying fractional amount of flux 
quantum \cite{Babaev:02} or nonmonotonic intervortex interactions originating 
in the additional length scales associated with the extra condensates; 
for a review see \cite{Babaev.Carlstrom.ea:12}.

Multicomponent models with biquadratic density-density interaction are
discussed, for example, in the context of superconductors with pair density 
wave order \cite{Agterberg.Tsunetsugu:08,Berg.Fradkin.ea:09a}, or in the 
context of interface superconductors such as SrTiO$_3$/LaAlO$_3$ 
\cite{Agterberg.Babaev.ea:14}.
Here we investigate the properties of topological defects in two-component 
models, in an immiscible phase where there is strong biquadratic interaction 
between condensates that penalizes coexistence of both condensates. 
This is modelled by a field theory of a doublet of complex fields that 
have a $U(1)\times U(1)\times\mathbb{Z}_2$ symmetry. 
In the immiscible case that occurs for strong biquadratic interaction, 
the ground-state spontaneously breaks a $U(1)\times\mathbb{Z}_2$
part of the symmetry of the theory.

We show that despite the fact that only one condensate exists in the 
ground-state, the topological defects' physics is dramatically altered 
because of the existence of the suppressed condensate. 
Depending on the values of the symmetry-breaking term, two ground-state 
phases with different broken symmetries are found. In the first phase, 
the ground-state spontaneously breaks the $U(1)\times U(1)$
part of the symmetry and both components of the doublet are equal and nonzero. 
In the second phase, only one component assumes nonzero ground-state density 
and the ground-state spontaneously breaks $U(1)\times\mathbb{Z}_2$. 
There, the $U(1)$ part associated with the vanishing condensate, 
is unbroken. Here, we are principally interested in the latter phase, where 
the spectrum of topological excitations features both domain walls and vortices.

We demonstrate that in this phase, the coexistence of both kinds of 
topological defects gives interesting defects that are vortices, 
but comprising a domain wall as well. That is, it resembles asymptotically 
a vortex carrying multiple flux quanta, but the magnetic field is 
localized along a circular domain wall at a given distance from the center. 
The overall object looks like a pipe. We thus refer to these configurations 
as pipelike vortices, in analogy with the discussion of pipelike vortices
in current-carrying two-component condensates \cite{Chernodub.Nedelin:10}.

Below, we introduce the simple two-component Ginzburg-Landau model that 
has $U(1)\times U(1)\times\mathbb{Z}_2$ symmetry. 
We then characterize the different possible ground-state phases of that 
model. Finally, we numerically investigate the properties of vortices within 
the phase where both components cannot coexist. We demonstrate that 
there exists a regime, where pipelike vortices form and they are stable. 
We eventually discuss the response to an external applied magnetic field.


The model considered here is a Ginzburg-Landau model of two charged 
condensates described by two complex fields $\psi_1$ and $\psi_2$.
These can be cast into a single complex vector $\Psi$, as 
$\Psi^\dagger=(\psi_1^*,\psi_2^*)$. The theory can be written 
as a theory with a global $\groupSU{2}$ symmetry that is explicitly 
broken by an extra inter-component term:
\Equation{FreeEnergy}{
 \mathcal{F}= \frac{\B^2}{2}+\frac{1}{2}|\D\Psi|^2   	
+\frac{\Lambda}{2}\Big(\Psi^\dagger\Psi-\Psi_0^2\Big)^2
+\delta|\psi_1|^2|\psi_2|^2
\,.
}
In addition to the coupling to the vector potential $\A$ of the 
magnetic field, through the kinetic term $\D\equiv\Grad+ie\A$ 
[and $|\D\Psi|^2 :=(\D\Psi)^\dagger\D\Psi$], the two condensates 
interact through the inter-component biquadratic density interaction 
$(\Lambda+\delta)|\psi_1|^2|\psi_2|^2$.
The theory is thus invariant under local $\groupU{1}$ transformations 
$\A\to\A-\Grad\chi$ and $\psi_a\to\Exp{i\chi}\psi_a$, for arbitrary $\chi(\x)$.
The potential has an $\groupSU{2}$ symmetry that is explicitly broken 
by the last term when $\delta\neq0$. When $\delta=0$, the theory is 
sometimes called semilocal $\groupSU{2}\times\groupU{1}$ since it has 
both a global $\groupSU{2}$ symmetry and a local $\groupU{1}$ symmetry 
group
\cite{
[{This theory was investigated in the framework of high energy physics, 
where such a symmetry group may occur as a sector of supersymmetric field
theories or in the context of grand unification theories. Note that it also 
corresponds to the bosonic sector of Weinberg-Salam theory where 
the non-Abelian gauge field decouples, for a review see: }] 
[{}] Achucarro.Vachaspati:00}.
For generic values of $\delta$, the symmetry of the theory is 
$\groupU{1}\times\groupU{1}\times\groupZ{2}$. That is, each $\groupU{1}$ is 
associated with independent global phase rotation of a condensate $\psi_a$, 
while the $\groupZ{2}$ symmetry is associated to the invariance under the 
discrete operation that permutes both condensates $\psi_1\leftrightarrow\psi_2$.

\begin{figure}[!htb]
\hbox to \linewidth{ \hss
\includegraphics[width=.75\linewidth]{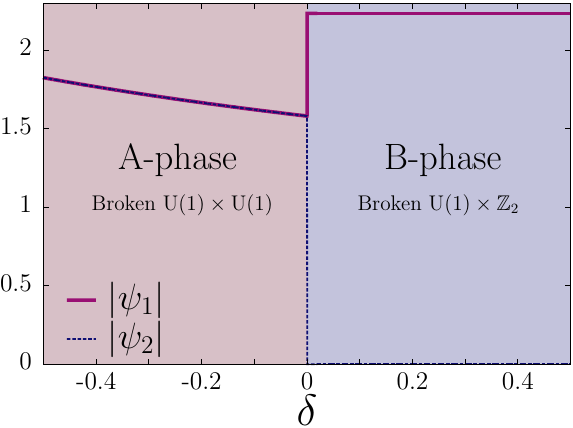}
\hss}
\vspace{-0.3cm}
\caption{
(Color online) -- 
Ground-state properties of the model. This shows the ground-state densities 
as functions of the explicit symmetry-breaking parameter $\delta$. The 
parameters are $\Psi_0^2=5$, $\Lambda=1$ and $e=0.8$. 
%
While changing the control parameter $\delta$, the system undergoes 
a phase transition at $\delta=0$. For negative $\delta$, the A-phase, 
both condensates have equal nonzero density and the ground-state breaks 
$\groupU{1}\times\groupU{1}$ symmetry. In the B-phase, for positive values 
of $\delta$, the repulsion between condensates is strong enough to penalize 
co-existence of both condensates and only one component has nonzero ground 
state density. The broken symmetry is thus $\groupU{1}\times\groupZ{2}$.
}
\label{Fig:PhaseDiag}
\end{figure}

The $\groupU{1}\times\groupU{1}\times\groupZ{2}$ symmetry of the 
theory is spontaneously broken by the ground-state. Depending on the 
symmetry-breaking parameter $\delta$, there are two different phases 
of the model Eq.~\Eqref{FreeEnergy}. 
When $\delta<0$, both condensates have the same density 
$|\psi_1|=|\psi_2|=\Lambda\Psi_0/(\delta+2\Lambda)$. We call this regime 
the A-phase, or miscible regime. In the B-phase, the immiscible regime 
we are principally interested in, only one condensate has nonzero density 
$(|\psi_1|,|\psi_2|)=(\Psi_0,0)$ or $(0,\Psi_0)$ and it is stable when 
$\delta>0$. This is summarized in \Figref{Fig:PhaseDiag}.
In the higher symmetry state where the $\groupSU{2}$ symmetry is not 
explicitly broken ($\delta=0$), only the total density is fixed, 
$|\psi_1|^2+|\psi_2|^2=\Psi_0^2$, and there is a continuous degeneracy 
to choose the relative density.

Symmetrywise both phases are different. In the A-phase, the ground-state 
breaks the $\groupU{1}\times\groupU{1}$ symmetry and since it is invariant 
under permutation of $\psi_1$ and $\psi_2$, it has an unbroken $\groupZ{2}$ 
symmetry. On the other hand, the B-phase spontaneously breaks the 
$\groupU{1}\times\groupZ{2}$ part. The unbroken $\groupU{1}$ symmetry 
is associated to the condensate that has zero density. There, the 
topological defects associated with the broken $\groupZ{2}$ symmetry 
are domain walls interpolating between the two inequivalent ground 
states $(\Psi_0,0)$ and $(0,\Psi_0)$. On the other hand, vortices are 
the topological defects associated with broken $\groupU{1}$ symmetries.
In two spatial dimensions, closed domain walls are topologically trivial 
and thus collapse for dynamical reasons. We show below that interaction 
with vortices can change that behavior.


\begin{figure}[!htb]
\hbox to \linewidth{ \hss
\includegraphics[width=\linewidth]{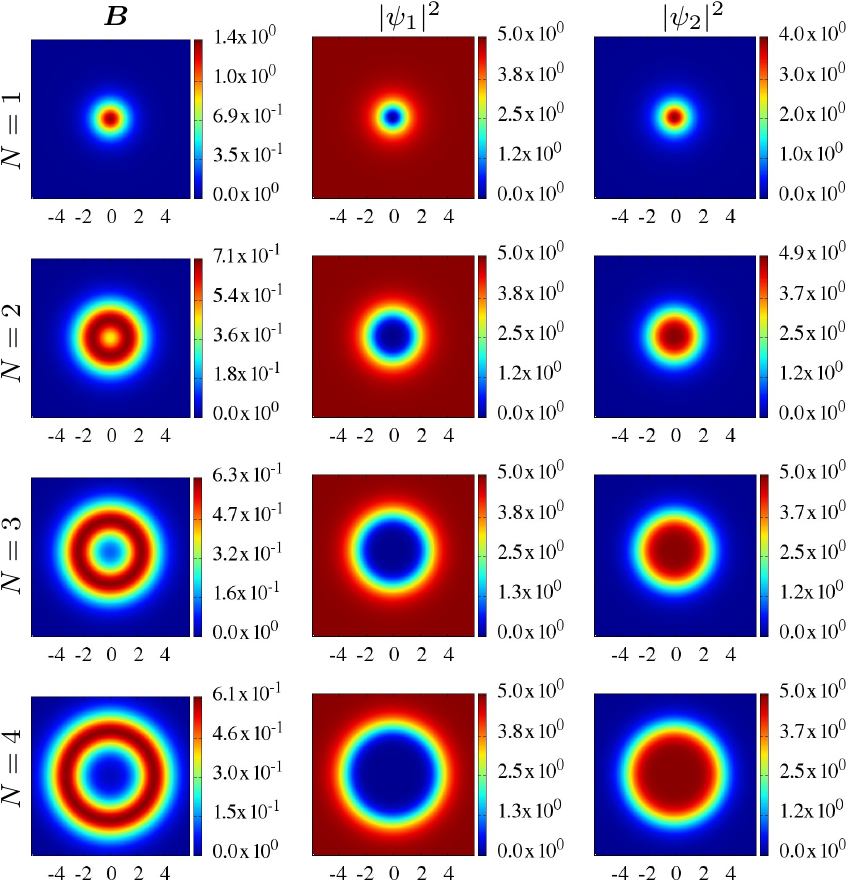}
\hss}
\vspace{-0.3cm}
\caption{
(Color online) -- 
Vortex solutions in the B-phase of \Figref{Fig:PhaseDiag}, for the 
symmetry-breaking parameter $\delta=0.02$. The first column shows 
the magnetic field and the second and third column, respectively show 
the densities $|\psi_1|^2$ and $|\psi_2|^2$. The different lines show 
configurations with different vorticity $N=1$, $2$, $3$ and $4$ respectively. 
In the B-phase, only one condensate has nonzero ground-state density. 
Here this is $\psi_1$, while $\psi_2$ vanishes asymptotically.
Despite not being in a type-1 regime, objects carrying multiple flux quanta 
are formed. The essential difference from type-1 vortices is that here
the magnetic field vanishes at the vortex center.
}
\label{Fig:Pipes}
\end{figure}

We consider field configurations varying in the $xy$ plane and assume 
invariance with respect to translations along the $z$ axis. To investigate 
the properties of topological defects, we numerically minimize the free 
energy [Eq.~\Eqref{FreeEnergy}] within a finite element framework \cite{Hecht:12}. 
That is, for a given choice of parameters, a starting configuration with 
the desired winding is created and the energy is minimized with a non-linear 
conjugate gradient algorithm. 
\cite{
[{For detailed discussion on the numerical methods, see for example 
appendix in:} ] [{}] Garaud.Babaev:14a}
The results of these simulations for vortices in the B-phase, for 
small $\delta$, are shown in \Figref{Fig:Pipes}. First, consider a 
configuration carrying a single flux quantum. There, the component 
that has nonzero ground-state density ($\psi_1$) forms a vortex. 
At the core of $\psi_1$, because there is less density, it becomes 
beneficial to give a nonzero value to $\psi_2$. The suppressed 
condensate $\psi_2$ condenses in the vortex core. The corresponding 
interface energy is positive, so it is preferable to minimize it. 

\begin{figure}[!htb]
\hbox to \linewidth{ \hss
\includegraphics[width=\linewidth]{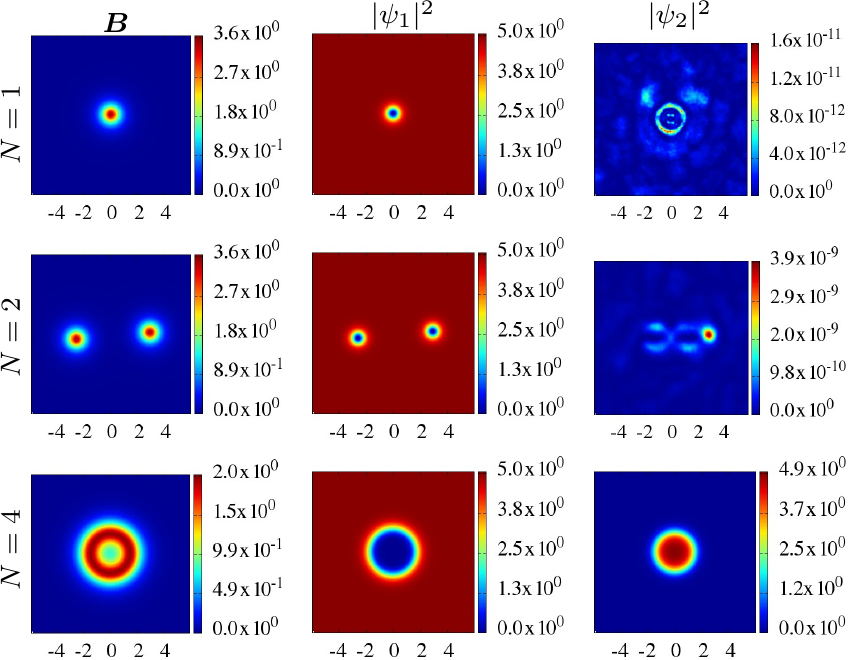}
\hss}
\vspace{-0.3cm}
\caption{
(Color online) -- 
Deep into the B-phase, vortex structures are substantially different from 
those obtained for small $\delta$. Here, displayed quantities are the same 
as in \Figref{Fig:Pipes} and the symmetry-breaking parameter is $\delta=0.4$.
It is no more beneficial to condense $\psi_2$ inside the vortex core. 
As a results, vortices are simply ordinary vortex solutions embedded in the 
$\groupU{1}\times\groupU{1}\times\groupZ{2}$ theory [Eq.~\Eqref{FreeEnergy}].
Nevertheless, pipelike vortices may be constructed here but they are metastable. 
That is, they still can exist but only as local minima of the energy functional. 
We find that typically configurations carrying a small number of flux quanta 
easily decay into vortices during the relaxation process, as shown on the 
second line. Solutions with larger $N$ form but their energy is larger than 
the one of $N$ isolated vortices. Note that the energy difference becomes 
smaller as the number flux quanta increase.
}
\label{Fig:MetaStable}
\end{figure}

For multiple quanta configurations, there is a competition between the 
type-2-like repulsion originating in vortices of $\psi_1$ and the 
attraction to minimize the interface energy of $\psi_2$ that condenses 
in the core. For small $\delta$, it is always preferred to form a bound 
state of vortices in order to minimize the interface energy. It results 
in a circular domain at the center of which $\psi_1=0$ and the condensate 
$\psi_2=\Psi_0$. At a certain distance, depending on the number of 
enclosed flux quanta, $\psi_1$ recovers its ground-state density while 
$\psi_2$ is completely suppressed. Thus there is a circular domain wall 
while outgoing from the vortex center. Since the ground-state is realized 
both at the center and asymptotically, the magnetic field is screened 
everywhere except at the domain wall. The overall configuration looks 
like a pipe, thus we refer to this as a pipelike vortex. Pipelike 
vortices are different from type-1 multiquanta vortices for which the 
magnetic field is non zero at the center.
The pipelike vortices we find here somehow recall pipelike solutions 
found for vortex configurations carrying persistent longitudinal current, 
in the $\delta=0$ case where the model has a global $\groupSU{2}$ symmetry
\cite{Chernodub.Nedelin:10}.
The reason for screening and the flux localization along the pipe is 
further discussed later in the paper.

At this point, it is important to recall that the B-phase has domain 
wall excitations that interpolate between  $(\Psi_0,0)$ and $(0,\Psi_0)$. 
Thus the interface previously mentioned is exactly such a domain wall.
For small $\delta$, domain walls have very small energy and their energy 
increases while going deeper in the B-phase. There, we can expect a change 
of behaviour because domain walls are more energetic and condensation 
in the vortex core becomes much smaller.
We find that, indeed, deeper in the B-phase, isolated vortices becomes 
preferred to pipelike vortices. Nevertheless, we do find the pipelike 
solutions despite the fact that isolated vortices with no condensation of 
$\psi_2$ in the core are preferred. Namely, 
we found configurations carrying $N$ flux quanta whose energy $E(N)$ is 
larger than the one of $N$ isolated vortices: $E(N)>NE(N=1)$. Such 
configurations are thus local minima of the energy functional and they 
differ by a few percent from isolated vortices. Such a metastable state 
is shown in \Figref{Fig:MetaStable}.


In substantially strong external field, vortex matter usually forms dense 
lattices. The previous results for isolated vortices inform us about the 
low-field physics. As discussed above, vortex configurations show very 
interesting structure comprising between the two kinds of topological defects 
that the theory allows. This may result in quite unusual properties of the 
solutions in high field. 
To investigate this, we simulate the response of the system to an external field 
${\bs H}=H_z\ez$ perpendicular to the plane. For this, the Gibbs free energy 
$\G=\F-\B\cdot{\bs H}$ is minimized, requiring that $\Curl\A={\bs H}$ 
on the boundary \cite{Garaud.Babaev:14a}. 
Note that since it is a finite sample with boundary Meissner currents the 
total flux through the sample does not have to be quantized, even in the 
standard vortex state.

\begin{figure}[!htb]
\hbox to \linewidth{ \hss
\includegraphics[width=\linewidth]{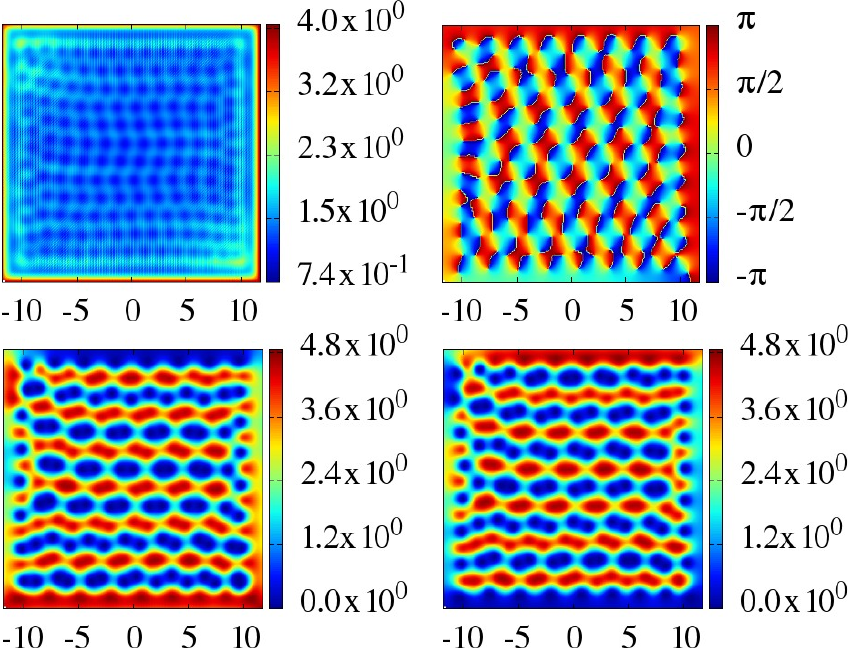}
\hss}
\vspace{-0.3cm}
\caption{
(Color online) -- 
Solution in an external field in the very special case of an $\groupSU{2}$ 
symmetric potential ($\delta=0$). The other parameters are the 
same as in \Figref{Fig:Pipes}. The panels on the first row display the 
magnetic field and the phase difference $\varphi_{12}=\varphi_2-\varphi_1$. 
The second line shows the densities $|\psi_1|^2$ and $|\psi_2|^2$, respectively. 
Here, although isolated vortices are unstable, they 
nonetheless form in the external field.
}
\label{Fig:Applied-SU2}
\end{figure}
We start by considering solutions in external field, at the boundary 
between A- and B- phases. At the point $\delta=0$ of the phase diagram, 
where the theory has the $\groupSU{2}$ symmetry, there are no stable 
type-2 vortices \cite{Hindmarsh:92,Hindmarsh:93,Achucarro.Vachaspati:00}.
Note that these vortices can be stabilized by having a twist of the phase 
in the $z$ direction \cite{Forgacs.Reuillon.ea:06,Forgacs.Reuillon.ea:06a}. 
This is somehow akin to having a symmetry-breaking term in the potential. 
Only short pieces of such twisted vortices are stable as they develop an 
unstable mode similar to hydrostatic Plateau-Rayleigh instability 
\cite{Garaud.Volkov:08}.
There are no stable isolated vortices in the theory with $\groupSU{2}$ symmetry, 
does not imply that its response in external field is trivial. Indeed, isolated 
vortices exhibit the spreading instability in the $\groupSU{2}$ case. However, 
there is additional constraint in external field. In \Figref{Fig:Applied-SU2}, 
we show that indeed the magnetic response is non-trivial. There are lines 
of vortices in a given condensate alternating with lines of vortices in the 
other one. Within a line vortices pair and form some kind of dimer. 
Note that finite-size effects and interaction with Meissner currents here 
play some role in deforming the lattice structure. That is, a perfect lattice 
can form only for tuned domains with periodic boundary conditions. The 
results in \Figref{Fig:Applied-SU2} should be understood as a typical state 
which would form in experiment, in mesoscopic samples.

\begin{figure}[!htb]
\hbox to \linewidth{ \hss
\includegraphics[width=\linewidth]{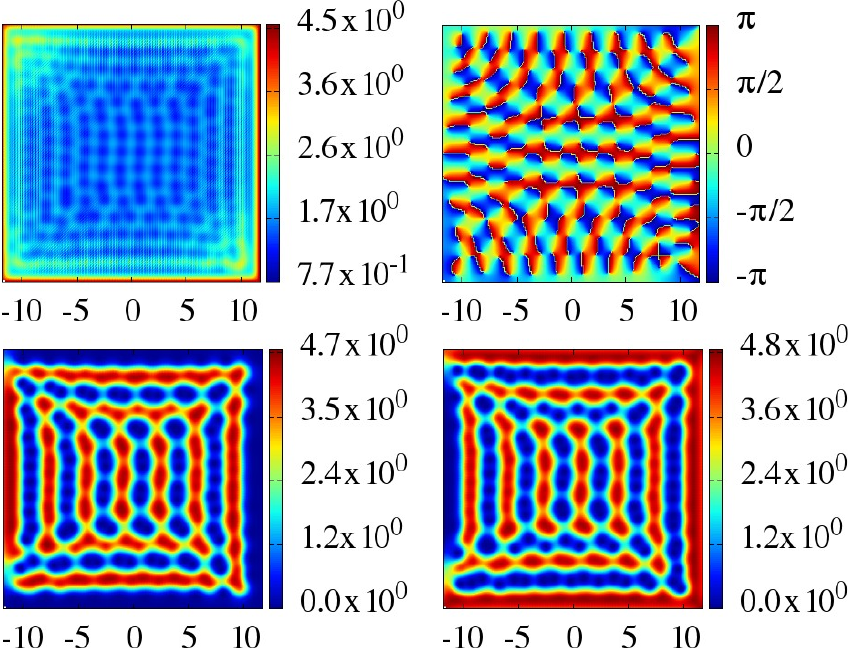}
\hss}
\vspace{-0.3cm}
\caption{
(Color online) -- 
Solution in an external field in the B-phase, for the same parameters 
as in \Figref{Fig:Pipes}. There the displayed quantities are the same 
as in \Figref{Fig:Applied-SU2}.
}
\label{Fig:Applied-Close}
\end{figure}
\begin{figure}[!hb]
\hbox to \linewidth{ \hss
\includegraphics[width=\linewidth]{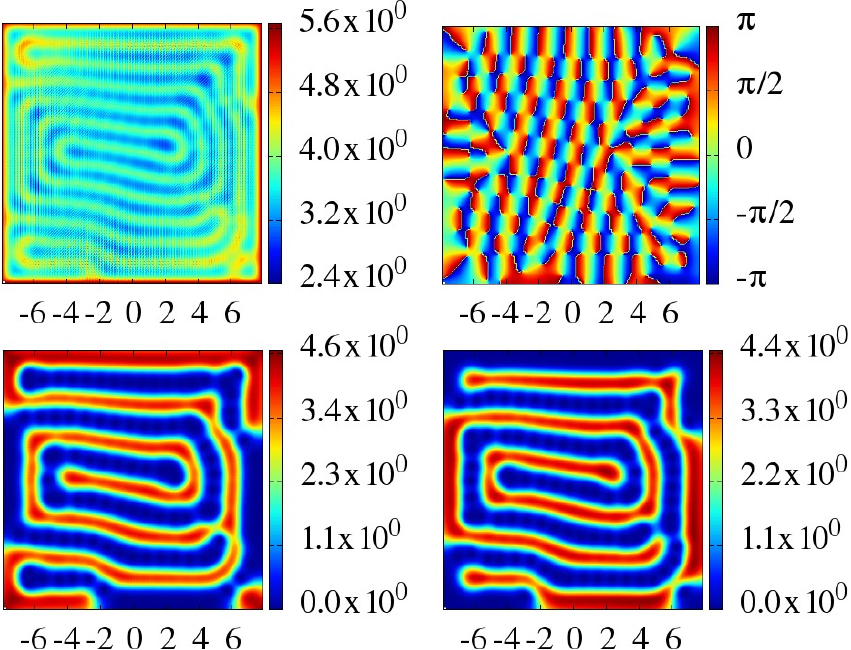}
\hss}
\vspace{-0.3cm}
\caption{
(Color online) -- 
Solution in an external field, the parameters are the same as in 
\Figref{Fig:Pipes}, but slightly deeper into the B-phase $\delta=0.2$. 
Displayed quantities are the same as in \Figref{Fig:Applied-SU2}. 
We see that there are alternating regions populated by vortices of 
different condensates. These alternating regions correspond (approximately) 
to the two inequivalent ground-states in the B-phase and thus they are 
separated by a domain wall that carries flux as in \Figref{Fig:Pipes}. 
The second panel showing the phase difference tells about position of 
singularity in both condensates.
}
\label{Fig:Applied-Deep}
\end{figure}

For small values of $\delta$, the behavior in external field combines the 
behaviour reported for the $\groupSU{2}$ symmetry in \Figref{Fig:Applied-SU2} 
and that deep into the B-phase shown in \Figref{Fig:Applied-Deep}. In this regime, 
displayed in \Figref{Fig:Applied-Close}, the dimers start to merge together.
Here again, finite-size effects and interaction with Meissner currents makes 
it very difficult to have a perfect lattice in this kind of simulation. 
It is interesting to note, that even when the stability properties of the 
topological excitations are completely different, the magnetic response 
can be quite similar to the one at the point with the $\groupSU{2}$ symmetry.

Deeper into the B-phase, the response to an external applied field starts 
to be quite different from those closer to the $\delta=0$ point. In 
\Figref{Fig:Applied-Deep}, we show that there are alternating regions 
populated by vortices of different condensates. These alternating regions 
correspond (approximately) to the two inequivalent ground-states in the B-phase. 
These regions are separated by a domain wall that carries flux as in 
\Figref{Fig:Pipes}. The domain walls form some kind of spiral covering the 
whole area of the sample. Such a pattern indicates that there is an important 
interplay between the two kinds of topological defects that the theory supports.

\begin{figure}[!htb]
\hbox to \linewidth{ \hss
\includegraphics[width=\linewidth]{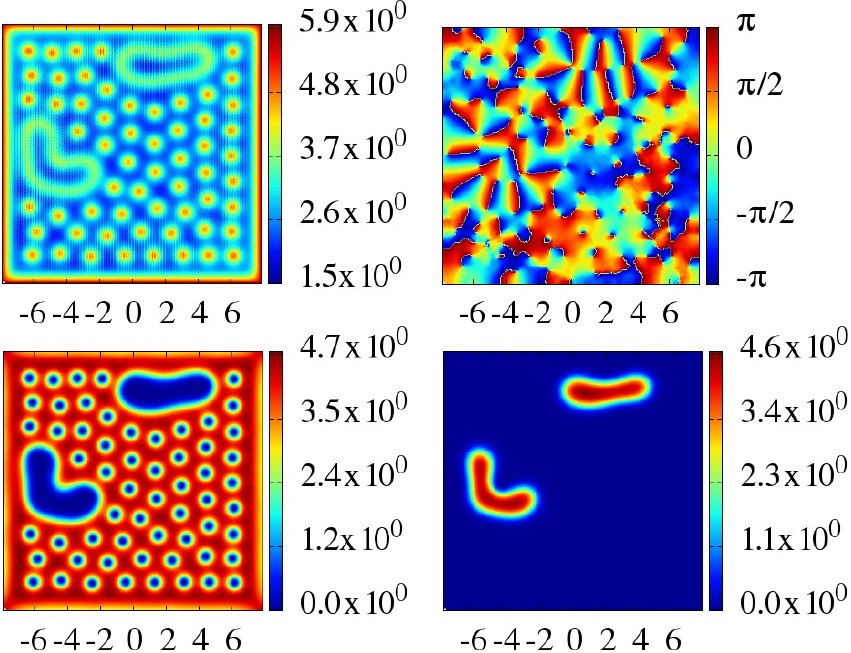}
\hss}
\vspace{-0.3cm}
\caption{
(Color online) -- 
Solution in an external field deeper in the in the B-phase than 
\Figref{Fig:Applied-Deep}, for values of the symmetry-breaking parameter 
$\delta=0.5$. There the displayed quantities are the same as in 
\Figref{Fig:Applied-Deep}.
Deep into the B-phase there 
is preference for usual vortices. However, pipelike 
vortices are metastable. Note that in external field the stability of 
pipelike vortices is somewhat enhanced by the pressure exerted by 
surrounding vortices.
}
\label{Fig:Applied-Deeper}
\end{figure}
In \Figref{Fig:MetaStable}, we showed that deep into the B-phase, 
isolated vortices are preferred to pipelike vortices. The latter 
may still exist but only as metastable states. The deeper into the 
B-phase, the pipelike vortices are less and less likely. However, 
as shown in \Figref{Fig:Applied-Deeper}, they still may coexist with 
regular vortices. Indeed, even if an isolated pipelike vortex is 
very sensitive to perturbations (as is the case deep into the B-phase), 
its stability is improved by the surrounding vortices that exert 
a `pressure' on the pipelike vortex, making its decay more 
difficult.

\begin{figure}[!htb]
\hbox to \linewidth{ \hss
\includegraphics[width=\linewidth]{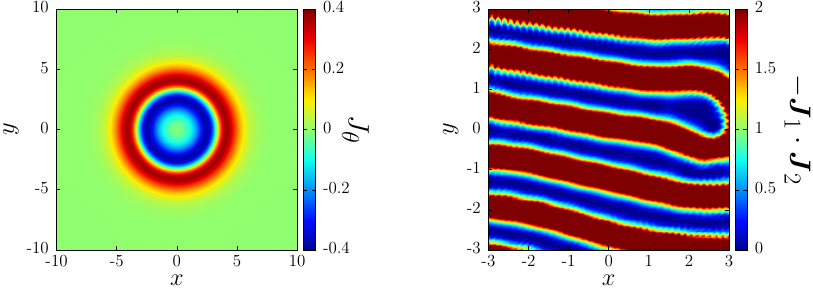}
\hss}
\vspace{-0.3cm}
\caption{
(Color online) -- 
Structure of superconducting currents. The first panel displays the total 
current for a pipelike vortex ($N=4$), from \Figref{Fig:Pipes}. Clearly, 
the supercurrent flows in opposite directions outside and inside the pipe. 
The region where the current changes its sign is region where the counterflow 
$-\J_1\cdot\J_2$ is maximum. In this region the magnetic field is high. 
The right panel shows a different example of stripe-like counterflow 
(zoomed in) for the simulation in external field shown, in 
\Figref{Fig:Applied-Deep}.
}
\label{Fig:Currents}
\end{figure}

The reason why the magnetic flux is localized along the domain wall 
can be understood as follows. From Amp\`ere's law $\Curl\B+\J=0$, 
the total supercurrent reads as $\J=e\Im(\Psi^\dagger\D\Psi)$ and the 
contribution due to each component is $\J_a=e\Im(\psi_a^*\D\psi_a)$.
For pipelike vortices, the supercurrents of both components flow in 
opposite directions. $J_2$ due to $\psi_2$ screens the magnetic field 
inside the domain, while $\psi_1$ is responsible for screening in the 
exterior. As a result, the only region where the magnetic flux penetrates 
is the domain wall. The structure of the superconducting currents in 
the pipelike vortices can be seen from \Figref{Fig:Currents}.


In this paper, we investigated the physical properties of topological 
defects in a Ginzburg-Landau model of a two-component superconductor that 
have density-density interaction. It can be seen as a model having an 
$\groupSU{2}$ potential supplemented by a term that explicitly breaks 
it down to $\groupU{1}\times\groupU{1}\times\groupZ{2}$. Depending on 
the symmetry-breaking parameter $\delta$, this model has two physically 
different phases. When $\delta>0$, the discrete $\groupZ{2}$ part of the 
symmetry is spontaneously broken, while when $\delta<0$ it is not. 
We have been focusing on the former phase where condensates cannot coexist 
and the ground-state is either $(\Psi_0,0)$ or $(0,\Psi_0)$. There, 
two kinds of topological defects are possible: domain walls and vortices.

We have shown that, for small symmetry-breaking term, vortices 
form bound states carrying multiple quanta of flux and exhibit  
properties of domain walls at the same time. The resulting 
configuration is a cylindrical inclusion of the component $\psi_2$ 
inside a whole domain where $\psi_1=\Psi_0$. The interface between both 
regions is a (cylindrical) domain wall where the flux is localized and 
that resembles a pipe.  
These pipelike vortices are stable only for small $\delta>0$. However, 
deeper in the phase separated regime, they can still exist, but they are 
metastable.

\begin{acknowledgments}

We acknowledge fruitful discussions with D. F. Agterberg.
This work is supported by the Knut and Alice Wallenberg Foundation 
through a Royal Swedish Academy of Sciences Fellowship, by the 
Swedish Research Council grants 642-2013-7837, 325-2009-7664 
and by NSF CAREER Award No. DMR-0955902.
The computations were performed on resources provided by the 
Swedish National Infrastructure for Computing (SNIC) at the National 
Supercomputer Center at Link\"oping, Sweden.
\end{acknowledgments}


\begin{thebibliography}{19}%
\makeatletter
\providecommand \@ifxundefined [1]{%
 \@ifx{#1\undefined}
}%
\providecommand \@ifnum [1]{%
 \ifnum #1\expandafter \@firstoftwo
 \else \expandafter \@secondoftwo
 \fi
}%
\providecommand \@ifx [1]{%
 \ifx #1\expandafter \@firstoftwo
 \else \expandafter \@secondoftwo
 \fi
}%
\providecommand \natexlab [1]{#1}%
\providecommand \enquote  [1]{``#1''}%
\providecommand \bibnamefont  [1]{#1}%
\providecommand \bibfnamefont [1]{#1}%
\providecommand \citenamefont [1]{#1}%
\providecommand \href@noop [0]{\@secondoftwo}%
\providecommand \href [0]{\begingroup \@sanitize@url \@href}%
\providecommand \@href[1]{\@@startlink{#1}\@@href}%
\providecommand \@@href[1]{\endgroup#1\@@endlink}%
\providecommand \@sanitize@url [0]{\catcode `\\12\catcode `\$12\catcode
  `\&12\catcode `\#12\catcode `\^12\catcode `\_12\catcode `\%12\relax}%
\providecommand \@@startlink[1]{}%
\providecommand \@@endlink[0]{}%
\providecommand \url  [0]{\begingroup\@sanitize@url \@url }%
\providecommand \@url [1]{\endgroup\@href {#1}{\urlprefix }}%
\providecommand \urlprefix  [0]{URL }%
\providecommand \Eprint [0]{\href }%
\providecommand \doibase [0]{http://dx.doi.org/}%
\providecommand \selectlanguage [0]{\@gobble}%
\providecommand \bibinfo  [0]{\@secondoftwo}%
\providecommand \bibfield  [0]{\@secondoftwo}%
\providecommand \translation [1]{[#1]}%
\providecommand \BibitemOpen [0]{}%
\providecommand \bibitemStop [0]{}%
\providecommand \bibitemNoStop [0]{.\EOS\space}%
\providecommand \EOS [0]{\spacefactor3000\relax}%
\providecommand \BibitemShut  [1]{\csname bibitem#1\endcsname}%
\let\auto@bib@innerbib\@empty
\bibitem [{\citenamefont {Nagamatsu}\ \emph {et~al.}(2001)\citenamefont
  {Nagamatsu}, \citenamefont {Nakagawa}, \citenamefont {Muranaka},
  \citenamefont {Zenitani},\ and\ \citenamefont
  {Akimitsu}}]{Nagamatsu.Nakagawa.ea:01}%
  \BibitemOpen
  \bibfield  {author} {\bibinfo {author} {\bibfnamefont {J.}~\bibnamefont
  {Nagamatsu}}, \bibinfo {author} {\bibfnamefont {N.}~\bibnamefont {Nakagawa}},
  \bibinfo {author} {\bibfnamefont {T.}~\bibnamefont {Muranaka}}, \bibinfo
  {author} {\bibfnamefont {Y.}~\bibnamefont {Zenitani}}, \ and\ \bibinfo
  {author} {\bibfnamefont {J.}~\bibnamefont {Akimitsu}},\ }\href {\doibase
  10.1038/35065039} {\bibfield  {journal} {\bibinfo  {journal} {Nature}\
  }\textbf {\bibinfo {volume} {410}},\ \bibinfo {pages} {63} (\bibinfo {year}
  {2001})}\BibitemShut {NoStop}%
\bibitem [{\citenamefont {Chu}\ \emph {et~al.}(2009)\citenamefont {Chu},
  \citenamefont {Koshelev}, \citenamefont {Kwok}, \citenamefont {Mazin},
  \citenamefont {Welp},\ and\ \citenamefont {Wen~(Eds.)}}]{Chu.Koshelev.ea:09}%
  \BibitemOpen
  \bibfield  {author} {\bibinfo {author} {\bibfnamefont {P.~C.}\ \bibnamefont
  {Chu}}, \bibinfo {author} {\bibfnamefont {A.}~\bibnamefont {Koshelev}},
  \bibinfo {author} {\bibfnamefont {W.}~\bibnamefont {Kwok}}, \bibinfo {author}
  {\bibfnamefont {I.}~\bibnamefont {Mazin}}, \bibinfo {author} {\bibfnamefont
  {U.}~\bibnamefont {Welp}}, \ and\ \bibinfo {author} {\bibfnamefont {H.-H.}\
  \bibnamefont {Wen~(Eds.)}},\ }\href {\doibase 10.1016/S0921-4534(09)00142-7}
  {\bibfield  {journal} {\bibinfo  {journal} {Physica C: Superconductivity}\
  }\textbf {\bibinfo {volume} {469}},\ \bibinfo {pages} {313} (\bibinfo {year}
  {2009})}\BibitemShut {NoStop}%
\bibitem [{\citenamefont {Maeno}\ \emph {et~al.}(1994)\citenamefont {Maeno},
  \citenamefont {Hashimoto}, \citenamefont {Yoshida}, \citenamefont
  {Nishizaki}, \citenamefont {Fujita}, \citenamefont {Bednorz},\ and\
  \citenamefont {Lichtenberg}}]{Maeno.Hashimoto.ea:94}%
  \BibitemOpen
  \bibfield  {author} {\bibinfo {author} {\bibfnamefont {Y.}~\bibnamefont
  {Maeno}}, \bibinfo {author} {\bibfnamefont {H.}~\bibnamefont {Hashimoto}},
  \bibinfo {author} {\bibfnamefont {K.}~\bibnamefont {Yoshida}}, \bibinfo
  {author} {\bibfnamefont {S.}~\bibnamefont {Nishizaki}}, \bibinfo {author}
  {\bibfnamefont {T.}~\bibnamefont {Fujita}}, \bibinfo {author} {\bibfnamefont
  {J.~G.}\ \bibnamefont {Bednorz}}, \ and\ \bibinfo {author} {\bibfnamefont
  {F.}~\bibnamefont {Lichtenberg}},\ }\href
  {http://dx.doi.org/10.1038/372532a0} {\bibfield  {journal} {\bibinfo
  {journal} {Nature}\ }\textbf {\bibinfo {volume} {372}},\ \bibinfo {pages}
  {532} (\bibinfo {year} {1994})}\BibitemShut {NoStop}%
\bibitem [{\citenamefont {Mackenzie}\ and\ \citenamefont
  {Maeno}(2003)}]{Mackenzie.Maeno:03}%
  \BibitemOpen
  \bibfield  {author} {\bibinfo {author} {\bibfnamefont {A.~P.}\ \bibnamefont
  {Mackenzie}}\ and\ \bibinfo {author} {\bibfnamefont {Y.}~\bibnamefont
  {Maeno}},\ }\href {\doibase 10.1103/RevModPhys.75.657} {\bibfield  {journal}
  {\bibinfo  {journal} {Rev. Mod. Phys.}\ }\textbf {\bibinfo {volume} {75}},\
  \bibinfo {pages} {657} (\bibinfo {year} {2003})}\BibitemShut {NoStop}%
\bibitem [{\citenamefont {Joynt}\ and\ \citenamefont
  {Taillefer}(2002)}]{Joynt.Taillefer:02}%
  \BibitemOpen
  \bibfield  {author} {\bibinfo {author} {\bibfnamefont {R.}~\bibnamefont
  {Joynt}}\ and\ \bibinfo {author} {\bibfnamefont {L.}~\bibnamefont
  {Taillefer}},\ }\href {\doibase 10.1103/RevModPhys.74.235} {\bibfield
  {journal} {\bibinfo  {journal} {Rev. Mod. Phys.}\ }\textbf {\bibinfo {volume}
  {74}},\ \bibinfo {pages} {235} (\bibinfo {year} {2002})}\BibitemShut
  {NoStop}%
\bibitem [{\citenamefont {Babaev}(2002)}]{Babaev:02}%
  \BibitemOpen
  \bibfield  {author} {\bibinfo {author} {\bibfnamefont {E.}~\bibnamefont
  {Babaev}},\ }\href {\doibase 10.1103/PhysRevLett.89.067001} {\bibfield
  {journal} {\bibinfo  {journal} {Phys. Rev. Lett.}\ }\textbf {\bibinfo
  {volume} {89}},\ \bibinfo {pages} {067001} (\bibinfo {year}
  {2002})}\BibitemShut {NoStop}%
\bibitem [{\citenamefont {Babaev}\ \emph {et~al.}(2012)\citenamefont {Babaev},
  \citenamefont {Carlstrom}, \citenamefont {Garaud}, \citenamefont {Silaev},\
  and\ \citenamefont {Speight}}]{Babaev.Carlstrom.ea:12}%
  \BibitemOpen
  \bibfield  {author} {\bibinfo {author} {\bibfnamefont {E.}~\bibnamefont
  {Babaev}}, \bibinfo {author} {\bibfnamefont {J.}~\bibnamefont {Carlstrom}},
  \bibinfo {author} {\bibfnamefont {J.}~\bibnamefont {Garaud}}, \bibinfo
  {author} {\bibfnamefont {M.}~\bibnamefont {Silaev}}, \ and\ \bibinfo {author}
  {\bibfnamefont {J.~M.}\ \bibnamefont {Speight}},\ }\href {\doibase
  10.1016/j.physc.2012.01.002} {\bibfield  {journal} {\bibinfo  {journal}
  {Physica C}\ }\textbf {\bibinfo {volume} {479}},\ \bibinfo {pages} {2}
  (\bibinfo {year} {2012})}\BibitemShut {NoStop}%
\bibitem [{\citenamefont {Agterberg}\ and\ \citenamefont
  {Tsunetsugu}(2008)}]{Agterberg.Tsunetsugu:08}%
  \BibitemOpen
  \bibfield  {author} {\bibinfo {author} {\bibfnamefont {D.~F.}\ \bibnamefont
  {Agterberg}}\ and\ \bibinfo {author} {\bibfnamefont {H.}~\bibnamefont
  {Tsunetsugu}},\ }\href {\doibase 10.1038/nphys999} {\bibfield  {journal}
  {\bibinfo  {journal} {Nature Physics}\ }\textbf {\bibinfo {volume} {4}},\
  \bibinfo {pages} {639} (\bibinfo {year} {2008})}\BibitemShut {NoStop}%
\bibitem [{\citenamefont {Berg}\ \emph {et~al.}(2009)\citenamefont {Berg},
  \citenamefont {Fradkin},\ and\ \citenamefont
  {Kivelson}}]{Berg.Fradkin.ea:09a}%
  \BibitemOpen
  \bibfield  {author} {\bibinfo {author} {\bibfnamefont {E.}~\bibnamefont
  {Berg}}, \bibinfo {author} {\bibfnamefont {E.}~\bibnamefont {Fradkin}}, \
  and\ \bibinfo {author} {\bibfnamefont {S.~A.}\ \bibnamefont {Kivelson}},\
  }\href {\doibase 10.1038/nphys1389} {\bibfield  {journal} {\bibinfo
  {journal} {Nat Phys}\ }\textbf {\bibinfo {volume} {5}},\ \bibinfo {pages}
  {830} (\bibinfo {year} {2009})}\BibitemShut {NoStop}%
\bibitem [{\citenamefont {Agterberg}\ \emph {et~al.}(2014)\citenamefont
  {Agterberg}, \citenamefont {Babaev},\ and\ \citenamefont
  {Garaud}}]{Agterberg.Babaev.ea:14}%
  \BibitemOpen
  \bibfield  {author} {\bibinfo {author} {\bibfnamefont {D.~F.}\ \bibnamefont
  {Agterberg}}, \bibinfo {author} {\bibfnamefont {E.}~\bibnamefont {Babaev}}, \
  and\ \bibinfo {author} {\bibfnamefont {J.}~\bibnamefont {Garaud}},\ }\href
  {\doibase 10.1103/PhysRevB.90.064509} {\bibfield  {journal} {\bibinfo
  {journal} {Phys. Rev. B}\ }\textbf {\bibinfo {volume} {90}},\ \bibinfo
  {pages} {064509} (\bibinfo {year} {2014})}\BibitemShut {NoStop}%
\bibitem [{\citenamefont {Chernodub}\ and\ \citenamefont
  {Nedelin}(2010)}]{Chernodub.Nedelin:10}%
  \BibitemOpen
  \bibfield  {author} {\bibinfo {author} {\bibfnamefont {M.~N.}\ \bibnamefont
  {Chernodub}}\ and\ \bibinfo {author} {\bibfnamefont {A.~S.}\ \bibnamefont
  {Nedelin}},\ }\href {\doibase 10.1103/PhysRevD.81.125022} {\bibfield
  {journal} {\bibinfo  {journal} {Phys. Rev.}\ }\textbf {\bibinfo {volume}
  {D81}},\ \bibinfo {pages} {125022} (\bibinfo {year} {2010})}\BibitemShut
  {NoStop}%
\bibitem [{\citenamefont {Achucarro}\ and\ \citenamefont
  {Vachaspati}(2000)}]{Achucarro.Vachaspati:00}%
  \BibitemOpen
  \bibfield  {author} {\bibinfo {author} {\bibfnamefont {A.}~\bibnamefont
  {Achucarro}}\ and\ \bibinfo {author} {\bibfnamefont {T.}~\bibnamefont
  {Vachaspati}},\ }\href {\doibase 10.1016/S0370-1573(99)00103-9} {\bibfield
  {journal} {\bibinfo  {journal} {Phys. Rept.}\ }\textbf {\bibinfo {volume}
  {327}},\ \bibinfo {pages} {347} (\bibinfo {year} {2000})}\BibitemShut
  {NoStop}%
\bibitem [{\citenamefont {Hecht}(2012)}]{Hecht:12}%
  \BibitemOpen
  \bibfield  {author} {\bibinfo {author} {\bibfnamefont {F.}~\bibnamefont
  {Hecht}},\ }\href {\doibase 10.1515/jnum-2012-0013} {\bibfield  {journal}
  {\bibinfo  {journal} {J. Numer. Math.}\ }\textbf {\bibinfo {volume} {20}},\
  \bibinfo {pages} {251} (\bibinfo {year} {2012})}\BibitemShut {NoStop}%
\bibitem [{\citenamefont {Garaud}\ and\ \citenamefont
  {Babaev}(2014)}]{Garaud.Babaev:14a}%
  \BibitemOpen
  \bibfield  {author} {\bibinfo {author} {\bibfnamefont {J.}~\bibnamefont
  {Garaud}}\ and\ \bibinfo {author} {\bibfnamefont {E.}~\bibnamefont
  {Babaev}},\ }\href {\doibase 10.1103/PhysRevB.89.214507} {\bibfield
  {journal} {\bibinfo  {journal} {Phys. Rev. B}\ }\textbf {\bibinfo {volume}
  {89}},\ \bibinfo {pages} {214507} (\bibinfo {year} {2014})}\BibitemShut
  {NoStop}%
\bibitem [{\citenamefont {Hindmarsh}(1992)}]{Hindmarsh:92}%
  \BibitemOpen
  \bibfield  {author} {\bibinfo {author} {\bibfnamefont {M.}~\bibnamefont
  {Hindmarsh}},\ }\href {\doibase 10.1103/PhysRevLett.68.1263} {\bibfield
  {journal} {\bibinfo  {journal} {Phys. Rev. Lett.}\ }\textbf {\bibinfo
  {volume} {68}},\ \bibinfo {pages} {1263} (\bibinfo {year}
  {1992})}\BibitemShut {NoStop}%
\bibitem [{\citenamefont {Hindmarsh}(1993)}]{Hindmarsh:93}%
  \BibitemOpen
  \bibfield  {author} {\bibinfo {author} {\bibfnamefont {M.}~\bibnamefont
  {Hindmarsh}},\ }\href {\doibase
  http://dx.doi.org/10.1016/0550-3213(93)90681-E} {\bibfield  {journal}
  {\bibinfo  {journal} {Nuclear Physics B}\ }\textbf {\bibinfo {volume}
  {392}},\ \bibinfo {pages} {461 } (\bibinfo {year} {1993})}\BibitemShut
  {NoStop}%
\bibitem [{\citenamefont {Forgacs}\ \emph
  {et~al.}(2006{\natexlab{a}})\citenamefont {Forgacs}, \citenamefont
  {Reuillon},\ and\ \citenamefont {Volkov}}]{Forgacs.Reuillon.ea:06}%
  \BibitemOpen
  \bibfield  {author} {\bibinfo {author} {\bibfnamefont {P.}~\bibnamefont
  {Forgacs}}, \bibinfo {author} {\bibfnamefont {S.}~\bibnamefont {Reuillon}}, \
  and\ \bibinfo {author} {\bibfnamefont {M.~S.}\ \bibnamefont {Volkov}},\
  }\href {\doibase 10.1103/PhysRevLett.96.041601} {\bibfield  {journal}
  {\bibinfo  {journal} {Phys. Rev. Lett.}\ }\textbf {\bibinfo {volume} {96}},\
  \bibinfo {pages} {041601} (\bibinfo {year} {2006}{\natexlab{a}})}\BibitemShut
  {NoStop}%
\bibitem [{\citenamefont {Forgacs}\ \emph
  {et~al.}(2006{\natexlab{b}})\citenamefont {Forgacs}, \citenamefont
  {Reuillon},\ and\ \citenamefont {Volkov}}]{Forgacs.Reuillon.ea:06a}%
  \BibitemOpen
  \bibfield  {author} {\bibinfo {author} {\bibfnamefont {P.}~\bibnamefont
  {Forgacs}}, \bibinfo {author} {\bibfnamefont {S.}~\bibnamefont {Reuillon}}, \
  and\ \bibinfo {author} {\bibfnamefont {M.~S.}\ \bibnamefont {Volkov}},\
  }\href {\doibase 10.1016/j.nuclphysb.2006.06.016} {\bibfield  {journal}
  {\bibinfo  {journal} {Nucl. Phys.}\ }\textbf {\bibinfo {volume} {B751}},\
  \bibinfo {pages} {390} (\bibinfo {year} {2006}{\natexlab{b}})}\BibitemShut
  {NoStop}%
\bibitem [{\citenamefont {Garaud}\ and\ \citenamefont
  {Volkov}(2008)}]{Garaud.Volkov:08}%
  \BibitemOpen
  \bibfield  {author} {\bibinfo {author} {\bibfnamefont {J.}~\bibnamefont
  {Garaud}}\ and\ \bibinfo {author} {\bibfnamefont {M.~S.}\ \bibnamefont
  {Volkov}},\ }\href {\doibase 10.1016/j.nuclphysb.2008.01.022} {\bibfield
  {journal} {\bibinfo  {journal} {Nucl. Phys.}\ }\textbf {\bibinfo {volume}
  {B799}},\ \bibinfo {pages} {430} (\bibinfo {year} {2008})}\BibitemShut
  {NoStop}%
\end{thebibliography}
%

\end{document}